\documentclass[fleqn,twoside]{article}
\usepackage[headings]{espcrc2}
\usepackage{amsthm,amssymb}

\readRCS
$Id: espcrc2.tex,v 1.2 2004/02/24 11:22:11 spepping Exp $
\usepackage{graphicx, balance}
\usepackage[figuresright]{rotating}

\newcommand{\AmS}{{\protect\the\textfont2
  A\kern-.1667em\lower.5ex\hbox{M}\kern-.125emS}}

\title{\textbf{Mathematical Model of Semantic Look - An Efficient Context Driven Search Engine}}

\author{Leena Giri G\address[DCSE]{Department of Computer Science and Engineering, University Visvesvaraya College of\\~Engineering, Bangalore University, Bangalore 560 001 India, Contact: leenagiri@gmail.com.\\},
{Srikanth P L\addressmark},
{S H Manjula \addressmark},
{K R Venugopal \addressmark},
L M Patnaik\address{Honorary Professor, Indian Institute of Science., Bangalore.}}

\runtitle{Mathematical Model of Semantic Look - An Efficient Context Driven Search Engine}
\runauthor{Leena Giri G, et al.,}

\begin{document}
\begin{abstract}
The World Wide Web (WWW) is a huge conservatory of web pages. Search Engines are 
key applications that fetch web pages for the user query. In the current generation web architecture, search engines treat keywords provided by the user as isolated keywords without considering the context of the user query. This results in a lot of unrelated pages or links being displayed to the user. Semantic Web is based on the current web with a revised framework to display a more precise result set as response to a user query. The current web pages need to be annotated by finding relevant meta data to be added to each of them, so that they become useful to Semantic Web search engines. Semantic Look explores the context of user query by processing the Semantic information recorded in the web pages. It is compared with an existing algorithm called OntoLook and it is shown that Semantic Look is a better optimized search engine by being more than twice as fast as OntoLook.  
\end{abstract}

\maketitle

\section{INTRODUCTION}
Semantic Web (Web 3.0) is the proliferation of unstructured documents of the web to a "web of data" [1]. In traditional web architecture there is less emphasis on  meta data of the web document during the data collection phase of the search engine and the concentration is more on classic approaches like Information Retrieval and Natural Language Processing. It is difficult to know the context or the role played by the web document designed for such approaches [2][3][4]. This is overcome by Semantic Web where enhanced version of meta data are embedded in the web pages as RDF [5] and Ontology [6]. Ontology defines the concepts and the relations between these concepts. RDF (Resource Description Framework) describes the web document in the form of triplets. Every RDF triplet is a composition of {\it subject}, {\it predicate} and {\it object}. {\it Subject} is an entity to be described, {\it object} is an entity which describes the subject and {\it predicate} is a relationship between {\it subject} and {\it object}; essentially every {\it predicate} describes the different context of the web page playing multiple roles. Both Ontologies and RDF are embedded in web pages forming the semantic annotation of a web page.  
\subsection{Motivation}
The existing search engines interpret the keywords of a user query in isolation without considering wholly, the context of the search query. Because of this, most of the results retrieved is irrelevant to the user query. This hits the performance and accuracy of search engines. The main purpose of providing multiple keywords is to make search based on a particular context. It is to say that nothing exists without context or relation. As an example, consider a scenario where a user has submitted the keywords "Ashoka+Bangalore+Hotel" with the intention to search for Hotel Ashoka in Bangalore. Traditional web search engines return all the web pages containing  the keywords "Ashoka, Bangalore and Hotel" without considering the context of the user query. Most of the web pages are irrelevant to the user query; where some pages may provide information on "Ashoka Pillar in Bangalore", some web pages may provide information on different hotels because of the keyword 'hotel', and some pages may provide information about Bangalore city because of the keyword 'Bangalore'.
\subsection{Contribution}
Semantic Look processes the semantic annotation embedded in web pages to perform the relevance analysis on the user query to retrieve the related URLs. It constructs the Ontology graphs which are eqvivalent to the Concept-Relation graph of OntoLook [7]. Semantic Look is optimized compared to OntoLook, where heavy weighted arcs are retained in every sub graph and only half of the number of least weighted arcs are pruned which mitigates the number of sub graphs to be processed. 
\subsection{Organization}
This Section describes the rest of the paper in brief. Section 2 describes a few earlier contributions to the implementation of Semantic Web based Search Engines. Section 3 defines the problem considered. Section 4 explains the architecture of Semantic Look. In Section 5 the mathematical model for Semantic Look is given by describing the notations used and the equations that result in a more accurate result set being displayed. Section 6 describes the algorithms required in the components of the system architecture. Section 7 gives the experimental results mentioning the data sets considered and a graphical representation of performance evaluation between OntoLook and Semantic Look. The paper concludes by mentioning the enhancement that can be incorporated in Semantic Look and the list of references considered by the authors.
\section{RELATED WORK}
Traditional search engines return keyword-isolated pages because of which many unrelated pages or web links are returned by them in response to the user's query. Semantic search engines can be categorized into different types of which context based search engine is one. This is the largest group and the aim here is to add semantic operations for better search results.  {\it Yufei Li et. al.,} [7] have proposed a prototype called OntoLook for performing relation based search to derive the context of user query. Concept-Relation Graphs (CRGs) are created with vertices representing concepts and edges representing the number of relations between these concepts. An algorithm generates sub graphs by pruning edges of the CRG irrespective of whether they are heavy weighted or less heavy weighted edges and this results in large number of sub graphs to be processed. 
\vskip 2mm 
 {\it Junghoo Cho, Hector Garcia-Molina, Lawrence Page} [8] have proposed the theory based on "Efficient Crawling Through URL Ordering", in order to obtain more "important" pages first. Obtaining important pages rapidly can be very useful when a crawler cannot visit the entire web in a reasonable amount of time. They define several important metrics, ordering schemes, and performance evaluation measures for this problem. They experimentally evaluate the ordering schemes on the Stanford University Web to prove that a crawler with a good ordering scheme can obtain important pages significantly faster than one without a good ordering scheme.
\vskip 2mm 
{\it Li Ding et. al.,} [9] have proposed the theory based on "Search on the Semantic Web". They propose this theory based on the fact that, in order to help human users and software agents find relevant knowledge on the Semantic Web, Swoogle, a search engine, discovers, indexes and analyzes the Ontologies and facts that are encoded in Semantic Web documents. {\it Natalya F Noy et. al.,} [10] have proposed the theory on creating "Semantic Web Contents". As researchers continue to create new languages in the hope of developing a Semantic Web, they still lack consensus on a standard. They describe how Protege-2000 - a tool for Ontology development and knowledge acquisition - can be adapted for editing models in different Semantic Web languages. Semantic Web is necessary to express information in a precise, machine interpretable form, so software agents processing the same set of data share an understanding of what the terms describing the data mean. 
\vskip 2mm
{\it Lastra J L M et. al.,} [11] study the use of Semantic Web Services in order to overcome this challenge. The use of Ontologies and explicit semantics enable performing logical reasoning to infer sufficient knowledge on the classification of processes that machines offer, and on how to execute and compose those processes to carry out manufacturing orchestration autonomously. {\it Yi Jin} [12] present an architecture of the Semantic Search Engine and our work shows how the fundamental elements of the Semantic Search engine can be used in the fundamental task of information retrieval. An improved version of the {\it tf-idf} (term frequency -inverse document frequency) algorithm is proposed to guarantee the retrieval of information resources in a more efficient by looking for items in which the the keywords that are searched for are more common than usual. 
\vskip 2mm
{\it Alexander Maedche et. al.,} [13] have presented an integrated enterprise-knowledge management architecture for implementing an Ontology based Knowledge Management System (OKMS) and have made a study on two critical issues related to working with Ontologies in real-world enterprise applications. {\it Wang Young-gui et. al.,} [14] have done analysis on application of Semantic Web to web mining and to build a semantic based web mining model under the framework of the Agent. The authors in [15] discuss clustering as a method to overcome the problem of searching through the list a search engine displays. The list that is displayed is extremely inconvenient to the users since it expects them to look into each page sequentially in an exhaustive manner which could result in relevant information being overlooked. {\it Mohammad Farhan Husain et. al.,} [16] discuss how current frameworks do not scale for storing large RDF graphs and describe a framework that is built using Hadoop by exploiting the cloud computing paradigm. 
\section{PROBLEM DEFINITION}
Given a set of keywords for a search, the main goal is to find the set of web pages related to the user search context by extracting the semantics behind the user query, where the result set contains most relevant web pages with unnecessary pages filtered from it. It is assumed that every web page consists of embedded RDF (e-RDF) and embedded Ontology (e-Ontology) forming the semantic annotation for a web page. The web pages with e-RDF and e-Ontology form the Semantic World Wide Web which is used by the crawler for crawling the e-RDF/e-Ontology in the web pages.
\vskip 2mm
Semantic Look, a variant of OntoLook, optimizes this search logic by retaining the high ranked edges in every sub graph, as they are relevant to the user query, and pruning only the least weighted arcs. As an example, if the number of edges in the CRG is 7 of which there are 4 less weighted arcs, OntoLook produces $2^7$ $i.e.$, 128 sub graphs. Semantic Look produces only 6 sub graphs, since it prunes half the number of least weighted arcs $i.e.$, $^4C_ 2$.
\section{SYSTEM ARCHITECTURE}
The Context Driven Search Engine includes Semantic Crawler, Semantic Parser, Semantic Look and Ontobase as components for drawing the context of user search, as shown in Figure 1. The context is drawn for a user query by extracting the relations among the keywords submitted by the user. The relations are recorded in RDF adhering to a particular Ontology, like the travel Ontology, and is embedded in every web page as semantic meta data. Semantic Web uses these meta data as information for searching the web pages.
\subsection{Semantic Crawler}
The Semantic Crawler collects the e-RDF/e-Ontology present in the web pages and invokes the corresponding parsers depending on the document type. Both RDF and Ontology are serialized as XML and embedded in a web page forming semantic meta data for a web page. The crawler performs the collection of web page contents and maps it to a web page database.
\begin{figure*}[ht!]
\center{\includegraphics[width=11cm]{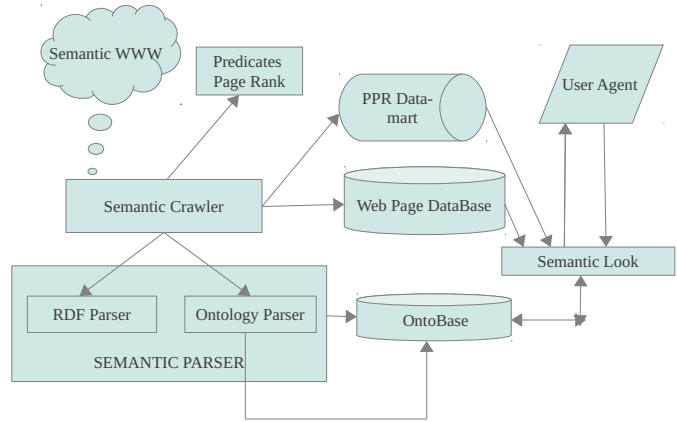}}
\caption{System Architecture}\label{Fig1}
\end{figure*}
\subsection{Semantic Parser}
This component of the search engine is responsible for parsing the incoming Semantic Annotation sent by the crawler. RDF parser parses the e-RDF documents to generate RDF-triplets and parses the e-Ontology to generate Ontology triplets. The generated triplets are mapped to Ontobase which is a knowledge base containing semantic information.
\subsection{Semantic Look}
The core component of the search engine is the Semantic Look. This utilizes the semantic information to develop all possible contexts for the user query. The context can be developed by extracting the relations between the keywords which are obtained by the corresponding Ontology and RDF triplets. The RDF triplets generated for the user query is used to fetch URL set.
\vskip 2mm
The Semantic Look\newline
\begin{itemize}
\item captures keywords and its corresponding concepts to form an Ontology graph where vertices represent concepts and edges represent relations between these concepts. Integers on the edges represent the number of relations between the concept pairs.
\item forms the less ranked arcs set and decide on the number of arcs to cut, by considering the average number of relations between the arcs.
\item cuts only the less ranked arcs to form the Ontology sub graph that will optimize the search result
\item processes the Ontology subgraph to produce all possible Ontology triplets from Ontobase
\item uses the Ontology triplets to form all possible RDF triplets for the user query
\item submits the RDF triplets to Ontobase to fetch URL set and sort according to the ranks assigned
\end{itemize}
The sequence of execution is as shown in Figure 2.
\begin{figure*}[ht!]
\center{\includegraphics[width=9cm]{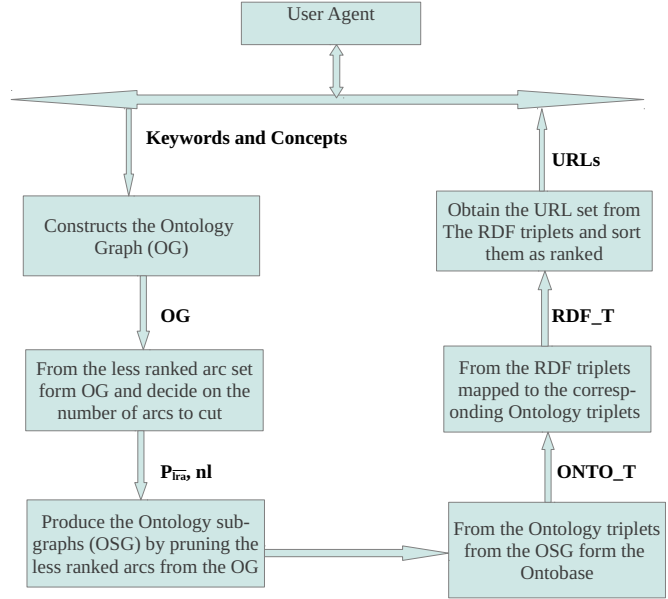}}
\caption{Functional Diagram}\label{Fig2}
\end{figure*}
\subsection{Ontobase}
Ontobase is a knowledge base containing semantic annotation of web pages which includes both Ontology and RDF triplets. Semantic Look uses this semantic annotation to obtain all possible RDF triplets defining the context of the user query.
\section{MATHEMATICAL MODEL}
This section describes the mathematical model  of Semantic Look using mathematical concepts and language. The model helps in explaining the system and to study the effects of different components, to analyse and predict its behaviour.
\begin{table}[ht!]
\begin{center}
\caption{Basic Notations}
\begin{tabular} {|l|l|}
\hline
\textbf{Notations}  &  \textbf{Meaning} \\\hline
K & Set of keywords submitted\\
         & by  the user. \\\hline
C & Set of concepts mapped for  \\
         &  the keywords given by the \\
         &  user \\\hline
$\lambda$ & Threshold: Minimum  \\
          &  support count for the occu-  \\
          &  rences of relations in a Web \\
          &  Page \\\hline
web  &  Set of URLs of web pages\\
        &  accessed according to \\
        & Web page index \\\hline
RDF\_T & Set of RDF triplets \\
           & accessed according to web \\
           & page index \\\hline
ONTO\_T & Set of Ontology triplets \\
            & indexed by Webpage index \\\hline
R & Set of relation vectors  \\
      & between the corresponding  \\
      & concept pairs given  by the \\
      & user \\\hline
OG & 2-D matrix representing \\
         & Concept-Relation graph \\\hline 
N & Total number of arcs in \\
   & Ontology graph \\\hline
nl & Total number of less ranked  \\
   & arcs where nl$<$n \\\hline
 p(r,w)  & Probability of occurrence \\
         & of relation r, R(w) in a\\
         & web page w. WI is referred  \\
 & to as {\it predicate frequency} \\\hline
KW &  Set of keyword vectors \\
       & accessed according to web \\
       & page index \\\hline
CW & Set of Concepts vectors \\
       & accessed according to web \\
       & page index \\\hline
R & Set of Relations vector  \\ 
 & between the  concept pairs  \\\hline
RW & Set of Relations vector \\\hline
\end{tabular}
\end{center}
\end{table}
Table 1 mentions the notations used by the authors and the relevant meanings of the notations used.
\subsection{Definitions}
{\bf  Webpage keywords}: It is the set of keywords vectors accessed according to web page index. It is given by $KW$=\{$KW_1$, $KW_2$...$KW_|web|$\}. Any $i^{th}$ vector is defined as
\begin{equation} 
KW_i = \{ k_l \mid 1 \leq l  \leq | KW_i  \mid  \}
\end{equation}
where $KW_i$ is the $i^{th}$ web page keyword and $k_l$ is the $l^{th}$ keyword in $KW_i$.
\\
{\bf Web page Concepts}: It is the set of concepts vectors accessed according to web page index. It is given by $CW$=\{$CW_1$, $CW_2$...$CW_|web|$ \}. Any $i^{th}$ vector is defined as 
\begin{equation}
CW_i=\{ c_l \mid  1 \leq l  \leq | CW_i \mid  \}
\end{equation}
where $CW_i$ is the $i^{th}$ vector of web page concepts mapped to the $i^{th}$ keyword vector, $i.e.$, $KW_i$ of web page keywords and $c_l$  is the $l^{th}$ concept in $CW_i$.
\\
{\bf Web page Predicates}: It is the set of relation vectors specifying the relations between the concept pairs in $CW$ and accessed according to webpage index. It is given by $RW$ $\leftarrow$\{$RW_1$, $RW_2$,...,$RW_|web|$\}. Any $i^{th}$ vector is defined as 
\begin{equation}
RW_i=\{r_{lk}, r_{kl} \mid 1 \leq l, k  \leq  C_2^{|CW_i|, l \ne k} \}
\end{equation}

where  $RW_i$ is the $i^{th}$vector of web page predicate with the domain and range concepts interchanged.
\\
{\bf Ontology Triplets of Web Page}: The Ontology triplet consists of subject concept called domain and object concept called range with predicate/relationship specyfying the relation between subject and object concepts. The Ontology triplets is a set on Ontology triplets vectors defined as  $ONTO\_T$=\{$ONTO\_T_1$, $ONTO\_T_2$,...,$ONTO\_T_|web|$\}. Any $i^{th}$ vector is defined as
\begin{equation}
ONTO\_T_i = \left \{
\begin{array} { l l }
{(CW_{id}, CW_{ij},CW_{ir})}\\
(CW_{ir},CW_{ij+1},CW_{id}) \mid \\
1 \le d,r \le C_2 \mid^{{CW}_i} \mid  \\
and \ 1 \le j \le \mid RW_i \mid \\
and \ d \ne r  \}
\end{array} \right \}
\end{equation}
where $CW_{id}$,$CW_{ir}$ are the domain and range concepts and $RW_{ij}$ is the $j^{th}$ relation in the $i^{th}$ web page specifying the relation between $CW_{id}$ and $CW_{ir}$. Similarly $RW_{ij+1}$ is a relation between $CW_{id}$ and $CW_{ir}$ with the domain and range concepts interchanged.
\\
{\bf RDF Triplets of Web Page}: The RDF triplet consists of subject and object with predicate/relationship specifying the relation between corresponding subject and object concepts. The RDF triplets is a set on RDF  triplets vectors defined as  $RDF\_T$=\{$RDF\_T_1$, $RDF\_T_2$,...,$RDF\_T_|web|$\}. Any $i^{th}$ vector is defined as
\begin{equation}
RDF\_T_i = \left \{
\begin{array} { l l }
(KW_{id}, RW_{ij},KW_{ir})\\
(KW_{ir},RW_{ij+1},KW_{id}) \mid \\
1 \le d,r \le C_2 \mid^{{KW}_i} \mid  \\
and \ 1 \le j \le \mid RW_i \mid \\
and \ d \ne r  \}
\end{array} \right \}
\end{equation}
where $KW_{id}$, $KW_{ir}$ are the domain and range concepts and $RW_{ij}$ is the $j^{th}$ relation in the $i^{th}$ web page specifying the relation between $KW_{id}$ and $KW_{ir}$. Similarly $RW_{ij+1}$ is a relation between $KW_{id}$ and $KW_{ir}$ with the domain and range concepts interchanged.
\vskip 2mm
{\bf Ontology Graph}: The concepts received from the user are paired to form Ontology graph, where the vertices represent the concepts and the edges represent the number of relationships existing between the concept pairs. The Ontology graph is represented as 2-D matrix where the value in $i^{th}$ row and $j^{th}$ column is defined as:
\begin{equation}
OG[i,j] = \left \{
\begin{array} { l l l }
\infty,\  if \  i \ne j \  and \mid R_{ij}\mid = 0\\
0 < \mid R_{ij} \mid  < \infty , \  if \  i \ne j \  \\
and \quad\mid R_{ij} \mid > 0 \\
0,\  if  \  i = j 
\end{array} \right \}
\end{equation}
where $R_{ij}$ $\in$  $R$  is a vector representing all the possible relations between the concept pairs $C_i$ and $C_j$ and is defined as:
\begin{equation}
R_{ij}  =  \left \{
RW_{wk}  \mid 
\begin{array} { l l }
1 \le w \le \mid\ RW_{w} \mid \\
1 \le d, r \le C_2^{\mid CW_{i}  \mid}, \\
1 \le k \le \mid RW_w \mid \\
and \ CW_{wd} = C_i  \\ 
and CW_{wr} = C_j \\
\end{array} \right \}
\end{equation}
\section{MODEL for SEMANTIC LOOK}
Semantic Look retains the high weighted arcs in the graph and prunes only the less weighted arcs to produce sub graphs from Ontology graph to be processed. The results are, therefore, relevant to the user query. The RDF triplets generated from the result of sub graph processing is submitted to web page database to fetch URL set.
\\
{\bf Theorem 1:}{\emph "The search engine time is reduced by pruning the less ranked arcs and the results are more relevant to user query".}\newline
{\bf Proof:} Let
\begin{equation}
{\emph  lra = \emph min(OG)}
\end{equation}
Find {\emph lra} which is the arc with the less weight. min(OG) returns minimum edge from Ontology graph. 
\begin{equation}
{\overrightarrow{P_{lra}}} =  \{ e_{ij} \mid 1 \le i \le OG \mid , lra = OG[i,j] \} 
\end{equation}
where \overrightarrow{P_{lra}} is a set of less ranked arcs and therefore $nl = \mid \overrightarrow{P_{lra}} \mid$ where $nl$ indicates number of such less ranked arcs and 
$nc=\lceil\frac{nl}{2}\rceil$ where {\it nc} indicates the number of less ranked arcs to be cut in Semantic Look. The average number of arcs are pruned from the Ontology graph. Since $nl<N$ and the total number of sub graphs which are candidates for processing are $2^{nl}<2^N$ and actual number of sub graphs processed are ${2^{nl}\choose{nc}}<{2^{N}\choose{nc}}$. Since high ranked arcs are retained in every sub graph, the result is more relevant to the user query and the search time is reduced by pruning only the less ranked arcs which produces $2^{nl}$ sub graphs which is less than $2^N$ sub graphs produced by not only pruning the less ranked arcs but also high ranked arcs from the Ontology graph.
\subsection{Implementation}
The proposed search engine called Semantic Look which is a variant of ONTOLOOK [1] optimizes the search engine time by pruning the less ranked arcs from the Ontology graph and retaining the high ranked arcs in every sub graph, which mitigates the number of sub graphs to be processed by the search engine.
The {\it Semantic Crawler} collects the semantic annotations embedded in every web page which includes embedded RDF and its corresponding Ontology (e-RDF/e-Ontology). The collected semantics and web page contents are stored in a database which is used by Semantic Look to perform the search for the user query. The {\it Semantic Parser} encapsulates the logic of the Ontology and RDF triplets to the database. 
\begin{table}[ht!]
\begin{center}
\caption{Algorithm for Semantic Crawler}
\begin{tabular}{|l|}
\hline
{\bf Input:} Semantic World Wide Web    \\                         
{\bf Output:} Web pages, e-RDF and e-Ontology\\
{\bf Process:} \\
{\emph web = Semantic World Wide Web} \\
{\bf for} each webpage {\emph w of web} \\
{\bf do} \\
\indent\hspace{.1in}{\it Li}=search$<$link$>$with type= \\
\indent\hspace{0.5in}”application/rdf+xml;” \\
{\bf for each} {\emph l $\in$ Li} \\
\indent\hspace{.1in}{\bf do} \\
\indent\hspace{.3in}{\emph O }= new Semantica Parser(); \\
\indent\hspace{.3in}{\emph Href} = href of $<$link$>$ \\
\indent\hspace{.3in}{\emph url} = url of {\it w} \\
\indent\hspace{.3in}{\emph rootTag} = parse the webpage whose\\
\indent\hspace{.3in}url is set in href of\\ 
\indent\hspace{.3in}$<$link$>$ to fetch the root element. \\
{\bf if}{\emph (rootTag =$<$ONTOLOGY$>$)} \\
\indent\hspace{.1in}{\bf then} \\
\indent\hspace{.2in}{\emph O} $\rightarrow$ {\emph OntologyParser(href,url)}; \\
\indent\hspace{.3in}{\bf else} \\
\indent\hspace{.4in}{\emph O} $\rightarrow$ {\emph RDFParser(href,url);} \\
\indent\hspace{.1in}{\bf done} \\
Store the contents of w in database. \\
{\bf done} \\
\hline
\end{tabular}
\end{center}
\end{table}

\begin{table}[ht!]
\begin{center}
\caption{Algorithm for Ontology Parser}
\begin{tabular}{|l|}
\hline
{\bf Input:} Ontology document URL (ourl), \\ 
 URL of Web Page(wurl) \\
{\bf Output:} Ontology Triplets mapped to  \\
  the database \\ 
{\bf Process:} \\
{\bf Step 1:} \\
{\emph i $=$ 0;} \\
{\bf for each}  {\emph ObjectProperty as op} \\
{\emph  and $++i$ $!=$ ObjectProperty.length }
{\bf do} \\
{\bf if}  (op.hasAttribute(“rdf:ID”)) {\bf then} \\
\indent\hspace{.2in}{\emph Relation} = {\emph sp.getAttribute(“rdf:ID”);} \\
{\bf for each} {\emph  op.childNodes as $ch$} {\bf do} \\
\indent\hspace{.2in}{\bf if} (ch.nodeName = “domain”)\\
\indent\hspace{.3in}{\emph domain} ={\emph  ch.getAttribute(“rdfs:resource”);} \\
\indent\hspace{.2in}{\bf else} \\
\indent\hspace{.3in}{\emph range} = {\emph ch.getAttribute(“rdfs:resource”);} \\
\indent\hspace{.2in}{\bf end if}  \\
{\bf done} \\
{\bf end if} \\
create \\
{$ONTO\_T_{wurli}$ = (Domain, Relation, Range);} \\
$ONTO\_T_{wurli}$ = $ONTO\_T_{wurl}$ $\cup$ \\
\indent\hspace{1.1in}$ONTO\_T_{wurli}$; \\
{\bf done} \\
$ONTO\_T$ = $ONTO\_T$ $\cup$ $ONTO\_T_{wurli}$; \\
repeat the above process for {\emph Data} \\
{\emph TypeProperty} and {\emph FunctionalProperty} \\
 insert Ontology triplet  to database. \\ 
\hline
\end{tabular} 
\end{center} 
\end{table}
\begin{table}[ht!]
\begin{center}
\caption{Algorithm for RDF Parser}
\begin{tabular}{|l|}
\hline
{\bf Input:} RDF document URL (ourl),\\  
URL of Web Page (wurl)  \\  
{\bf Output:} RDF\_TRIPLETS of wurl \\ 
 $i.e.$, RDF\_Twurl \\
{\bf Process:} \\
{\emph i} = {\emph 0};\\
\indent\hspace{.1in}{\bf for each} rdf:Description as {\it d} and \\
\indent\hspace{.2in}++i!=rdf:description.length \\
\indent\hspace{.2in}{\bf do}  \\
\indent\hspace{.3in}Subject = d.getAttribute(“rdf:about”); \\
\indent\hspace{.3in}{\bf for each} d.childNodes as {\it c} \\
\indent\hspace{.4in}{\bf do} \\
\indent\hspace{.4in}Relation=c.nodeName; \\
\indent\hspace{.4in}Object = c.getAttribute(“rdf:resource”); \\
\indent\hspace{.4in}{\bf done} \\
\indent\hspace{.3in}{\bf end for} \\
create RDF\_T-{wurl} U RDF\_T\_{wurli} \\
\indent\hspace{.1in}{\bf end for} \\
obtain corresponding ONTO\_T\_{wurl} \\
insert ONTO\_T\_{wurl} and RDF\_T\_{wurl} \\
to the database. \\
\hline
\end{tabular}
\end{center}
\end{table}

The {\it Semantic Crawler} crawls the web documents from the {\it Semantic World Wide Web} (Semantic-WWW) and invokes the {\it Ontology parser} if the collected document is the Ontology document or invokes the RDF-parser if the collected document is the RDF document. The crawler also stores the web pages in the database for future use as explained in the algorithm given in Table 2. The Ontology parser creates the Ontology triplets by fetching the domain and range concepts of {\it ObjectProperty, DatatypeProperty} or {\it FunctionalProperty} and maps the corresponding triplets to the database as given in the algorithm of Table 3. 
\vskip 2mm
The RDF parser looks for the {\it $<$rdf:descrip tion}$>$ element for forming the RDF-triplet and maps it to the corresponding Ontology triplets before it is added to the database. The {\it subjects/objects} are the instances of some Ontology concepts defined in the Ontology document which are described under the element called {\it $<$instances$>$} in RDF-document. The RDF-parser uses this information to map RDF-triplets to the Ontology triplets as explained in the algorithm of Table 4. 
{\it Semantic Look} dynamically constructs the Ontology graph from the concepts and corresponding keywords given by the user. The search engine cuts the less weighted arcs from the graph retaining high ranked arcs. There is no set criterion for deciding to cut such number of arcs but in this search engine average number of arcs are pruned from the Ontology graph to form the sub graphs. Since high ranked arcs are retained in every sub graph, the results obtained are relevant to the user query.
The sub graphs are processed to from all possible RDF-triplets which are submitted to the database to retrieve URL set. Since there is a probability that RDF-triplets are repeated in multiple web pages the final URL set is obtained by the intersection of URL sets of all RDF-triplets matching the user query as explained in the algorithm of Table 5. \\
\begin{table}
\begin{center}
\caption{Algorithm for Semantic Look}
\begin{tabular}{|l|}
\hline
{\bf Input:} Key words set $i.e.$, K{},\\
 Concepts set $i.e.$, C{} \\
{\bf Data Store:} Semantic World Wide Web \\
$i.e.$, Web{} \\
{\bf Output:} URL Set \\
{\bf Process:} \\
$Cn = \mid C \mid$ \\
lra = $ \infty $ \\
{\bf let} R = { R12, R21, R23, R32, \dots, Rcncn-1} \\
\indent\hspace{.1in}{\bf for each} i,j $<$ cn and i $\ne$ j  {\bf do}\\
\indent\hspace{.3in}$OG[i,j] = \mid R_{ij} \mid$ \\
\indent\hspace{.4in}{\bf if}  $(OG[i,j] < lra)$  {\bf then} \\
\indent\hspace{.5in} lra = OG[i,j] \\
\indent\hspace{.4in}{\bf end if} \\
\indent\hspace{.2in}{\bf done} \\
$\overrightarrow P_{lra} = \Phi $  \\
\indent\hspace{.1in}{\bf for each} i, j $<$ cn and i!=j {\bf do}\\
\indent\hspace{.3in}{\bf if} $(OG[i,j]= lra)$ {\bf then}\\
\indent\hspace{.4in} $\overrightarrow{P_{lra}} =\overrightarrow{P_{lra}} U e_{ij}$  \\
\indent\hspace{.3in}{\bf end if}
\indent\hspace{.2in}{\bf done}
$nl = \mid\overrightarrow{P_{lra}} \mid; ceil(\frac{nl}{2})$; \\
i =0; \\
Pstack; \\
{\bf While} ($i$ $<$ $nl$) {\bf do} \\
\indent\hspace{.2in}k=i \\
\indent\hspace{.2in}$.$push(\overrightarrow{P_{lra k}})  \\
\indent\hspace{.1in}{\bf while} (!Pstack.empty())  {\bf do}\\
\indent\hspace{.3in}{\bf if} (Pstack.length = nc)\\ 
\indent\hspace{.3in}OSG[ ]; \\
\indent\hspace{.1in}{\bf for each} $Pstack[j]$ as $e_{lm}$ {\bf do}\\
\indent\hspace{.3in}copy OG[ ] to OSG[ ] \\
\indent\hspace{.3in}[l,m] = OSG[m,l] = $\infty$ \\
\indent\hspace{.3in}Generate\_RDF\_T(OSG) \\
\indent\hspace{.2in}{\bf done} \\
\indent\hspace{.1in}Pstack$.$pop() \\ 
\indent\hspace{.1in}end if \\
{\bf else} \\
{\bf do} \\
\indent\hspace{.2in}{\bf if} (++k$\ne$nl) {\bf then}\\
\indent\hspace{.3in}$.$push(\overrightarrow{P_{lrak}}) \\
\indent\hspace{.2in}{\bf end if} \\
k = \overrightarrow{P_{lra}}$.$position(Pstack.top) \\
Pstack$.$pop() \\
{\bf done} \\
{\bf done} \\
{\bf done} \\
\hline 
\end{tabular} \\
\end{center}
\end{table}
\begin{tabular}{|l|}
\hline \\
/*Generate RDF Triplets to fetch URLs*/ \\
{\bf Procedure} {\it Generate\_RDF\_T(OSG)} \\
{\bf do} \\
{\bf URL} = $\phi$ \\
{\bf for} (i,$j<\mid OSG\mid$) \\
{\bf do} \\
$R_{ij} = RW_{wk}$ if \\
( 1 $<=$ w $<= \mid\ RW_{w} \mid$   and \\
\indent\hspace{.1in}1 $<=$ d, r $<= C_2^{\mid CW_{i} \mid}$, and \\
\indent\hspace{.1in}1 $<=$ k $<= \mid RW_w \mid$ \\
\indent\hspace{.2in}and $KW_{wd} = K_i$, $KW_{wr} = K_j$ )\\
{\bf for} $k< \mid R_{ij} \mid$ \\
\indent\hspace{.1in}{\bf do} \\
\indent\hspace{.2in}{\it urli} = $\{Web_i\mid(k_i,R_{ijk},k_j)$\\
\indent\hspace{.3in}$\in RDF\_T_i\}$ \\
\indent\hspace{.2in}{\it URL} = URL $\cap$ urli \\
\indent\hspace{.1in}{\bf done} \\
{\bf done} \\
rsort(url); \\
{\bf done} \\
\hline
\end{tabular} 
\section{EXPERIMENTAL RESULTS}
The dataset consists of 40 web pages with embedded 370 RDF and 85 Ontology triplets. The web pages are mapped to the web page database, where RDF and Ontology triplets are mapped to Ontobase. The RDF documents, Ontology documents and web pages form the Semantic World Wide Web. The XML serialized Ontology and RDF triplets are mapped to the tables in Ontobase which has complete information about the  triplets such as  subject, subject’s concept ,object, object’s concept, predicates and predicates type. The concept and predicates type are obtained from the Ontology document. Semantic Look is simulated on the context of the tourism portals. The search engine is generic but the e-Ontology and e-RDF is domain specific. The Semantic Crawler crawls the e-Ontology and e-RDF without any knowledge on the context where these documents are playing a role. The search context for a user query is established by extracting the relations among the supplied keyword. This is performed by {\it  Semantic Look}.
\vskip 2mm
The entire application is developed on {\it LAMPP} environment with{\it PHP} as underlying language for business logic. As shown in {\it Fig 3} the Semantic Look and Ontolook is compared with respect to the number of relations to be processed for different sets of keywords and concepts provided by the user.The difference in the number of sub graphs processed by OntoLook and Semantic Look is given in Table 7.

\begin{table*}[ht!]
\begin{center}
\caption{Sub Graphs Processed  for a Particular Combination Keywords and Relations}

\begin{tabular}{|c|c|c|c|c|c|}
\hline
\multicolumn{2} {|c|}{{No.of Keywords}} &
\multicolumn{2} {c|}{{No.of Relations}} &
\multicolumn{2} {c|}{{No.of Sgraphs processed}} \\
\hline
{OLook} & {SLook} &{OLook} & {SLook} & {OLook} & {SLook} \\ \hline
{8} & {8} & {25} & {10} & {5200300} & {252}  \\ \hline
{7} & {7} & {18} & {6}  & {48620}   & {20} \\ \hline
{5} & {5} & {9}  & {3}  & {26}      & {3} \\ \hline
{4} & {4} & {5}  & {2}  & {10}      & {2} \\ \hline
{3} & {3} & {3}  & {2}  & {3}       & {2} \\ \hline
\end{tabular} 
\end{center}
\end{table*}

Since in every sub graph high ranked edges are retained and only the selected less ranked edges are pruned, the number of sub graphs to be processed is less in Semantic Look compared to Ontolook. As shown in {\it Table 7}, the number of relations to be processed in Semantic Look is less than half of the number of relations processed by Ontolook as depicted in {\it Figure 3}.
\begin{figure}[ht!]
\centerline{\includegraphics [width=8cm]{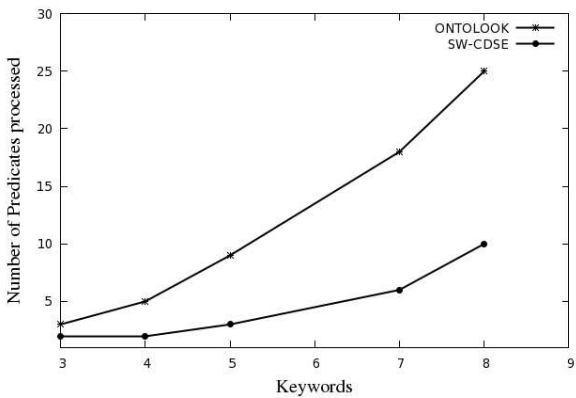}}
\caption{Keywords Predicates Processed}\label{Fig3}
\end{figure}
Every sub graph produces large number of duplicate RDF triplets which is submitted to the Ontobase to fetch URLs for every sub graph and intersection of these URL sets produce the distinct set of URLs as a result set for the user. The search time here includes the time for pruning the selected less ranked edges from the Ontology graph, producing the RDF triplets and database communication time for fetching the URLs set from it.  From {\it Table 7} and {\it Figure 3} it is shown that number of sub graphs produced in Semantic Look is less compared to Ontolook and therefore the number of RDF triplets produced in Semantic Look is less which in turn reduces the search time as compared with Ontolook. {\it Table 8} shows the number of RDF triplets processed and search time taken by Ontolook and Semantic Look. 
\begin{table*}[ht!]
\begin{center}
\caption{No. of RDF Triplets Produced and Search Time to process them for Combination Keywords and Relations}
\begin{tabular}{|c|c|c|c|c|c|}
\hline
\multicolumn{2} {|c|}{{Sub graphs processed}} &
\multicolumn{2} {c|}{{RDF triplets produced}} &
\multicolumn{2} {c|}{{Process Time}} \\ \hline
{OLook} & {SLook} & {OLook} & {SLook} & {OLook} & {SLook} \\ \hline
{5200300} & {252} & {701345778} & {81144} & {710039}    & {34$.$4076}\\ \hline
{48620}   & {20}  & {5209920}   & {4320}  & {21$.$0912} & {1$.$874}\\ \hline
{126}     & {3}   & {6832}      & {354}   & {3$.$2049}  & {0$.$216094} \\ \hline
{10}      & {2}   & {244}       & {120}   & {0$.$12911} & {0$.$0634} \\ \hline
{3}       & {2}   & {48}        & {46}    & {0$.$02599} & {$.$02399}\\ \hline
\end{tabular}
\end{center}
\end{table*}
\section{CONCLUSIONS}
Search engines in the current web architecture will not consider  the semantics  role  played by web pages in different context. The new generation of web  $i.e.$, Semantic  Web (web 3.0) considers  this  context  information  by  recording  the  semantic  information in the  form  of Ontologies and RDFs. A proof of concept called Semantic Look is proposed to produce relevant web pages by filtering unnecessary web documents from the result set.
\vskip 2mm
Semantic Look  extracts the semantics of the user query to know the context of user search. This work is based on the prototype called OntoLook which performs the exhaustive search of all the sub graphs of Ontology graph to produce URL set. Semantic Look is an optimized search engine compared to OntoLook which prunes less weighted edges from the OntoLook to produce less number of sub graphs for processing.
\vskip 2mm
Even though the number of sub graphs processed by Semantic Look is less as compared with OntoLook the number of RDF triplets produced will be huge and therefore in future work Semantic Look should be designed to run on the clusters of nodes using Map-Reduce Framework. Further optimization is achieved by running the crawler and pruning logic on the cluster. Since semantic information is embedded in the web page by the author and it is assumed to be true there is a chance of misleading the search engine by embedding false
semantic information. 
\small
\balance

\noindent{\includegraphics[width=1in,height=1.7in,clip,keepaspectratio]{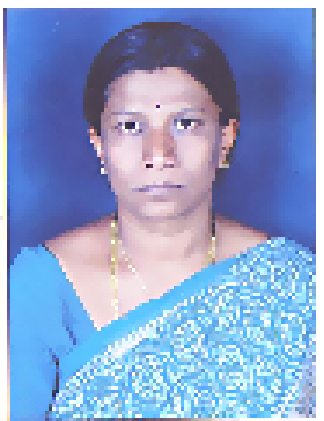}}
\begin{minipage}[b][1in][c]{1.8in}
{\centering{\bf {Leena Giri G}} is currently an Associate Professor in the Department of Computer Science, Dr. Ambedkar Institute of Technology, Bangalore. She obtained her Bachelor of Engineering from SJCE, Mysore. She received her M.Tech Degree  in Computer Science and Engi-}\\\\
\end{minipage}
neering from IIT Mumbai. Her research interest is in the area of Semantic Web.\\\\
\noindent{\includegraphics[width=1in,height=1.7in,clip,keepaspectratio]{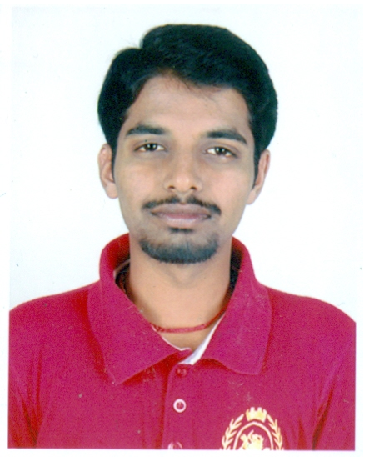}}
\begin{minipage}[b][1in][c]{1.8in}
{\centering{\bf {Srikanth P L}} received his Master's degree from the Department Computer Science and Engineering, University Visvesvaraya College of Engineering, Bangalore University, Bangalore. 
His research interest is in the area of Web Technology, Se-}\\  
\end{minipage}
mantic Web and Cloud Computing. \\\\
\noindent{\includegraphics[width=1in,height=1.7in,clip,keepaspectratio]{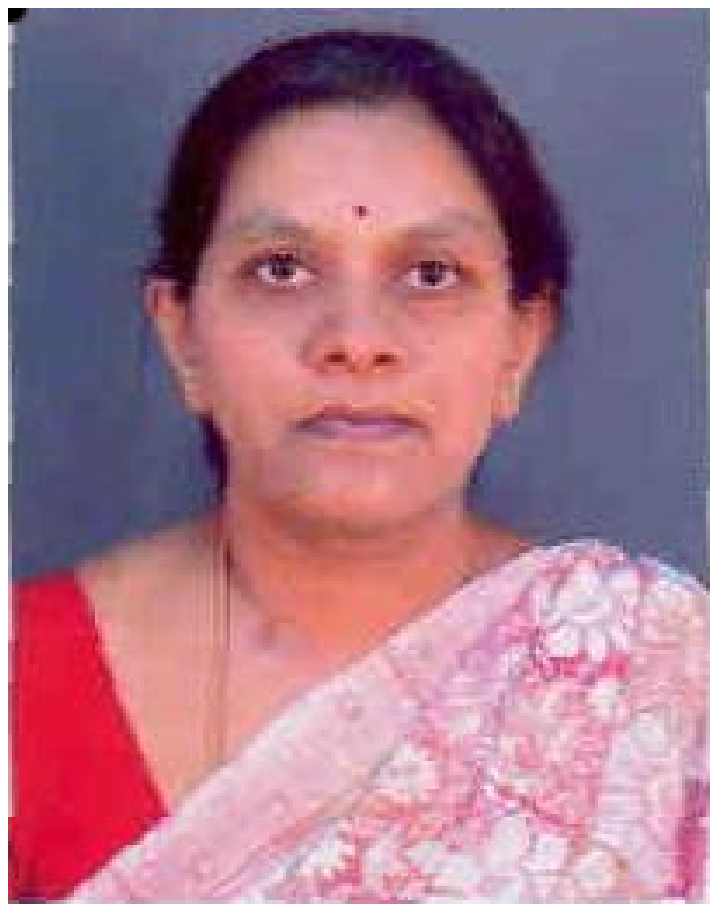}}
\begin{minipage}[b][1in][c]{1.8in}
{\centering{\bf {S H Manjula }} is currently the Chairman, Department of Computer Science and Engineering, University Visvesvaraya College of Engineering, Bangalore University, Bangalore. She obtained her Bachelor of Engineering and Masters Degree in Computer Science and Engineering from }\\\\  
\end{minipage}
University Visvesvaraya College of Engineering. She was awarded Ph.D in Computer Science from Dr. MGR University, Chennai. Her research interests are in the field of Wireless Sensor Networks and Data mining. \\\\
\noindent{\includegraphics[width=1in,height=1.7in,clip,keepaspectratio]{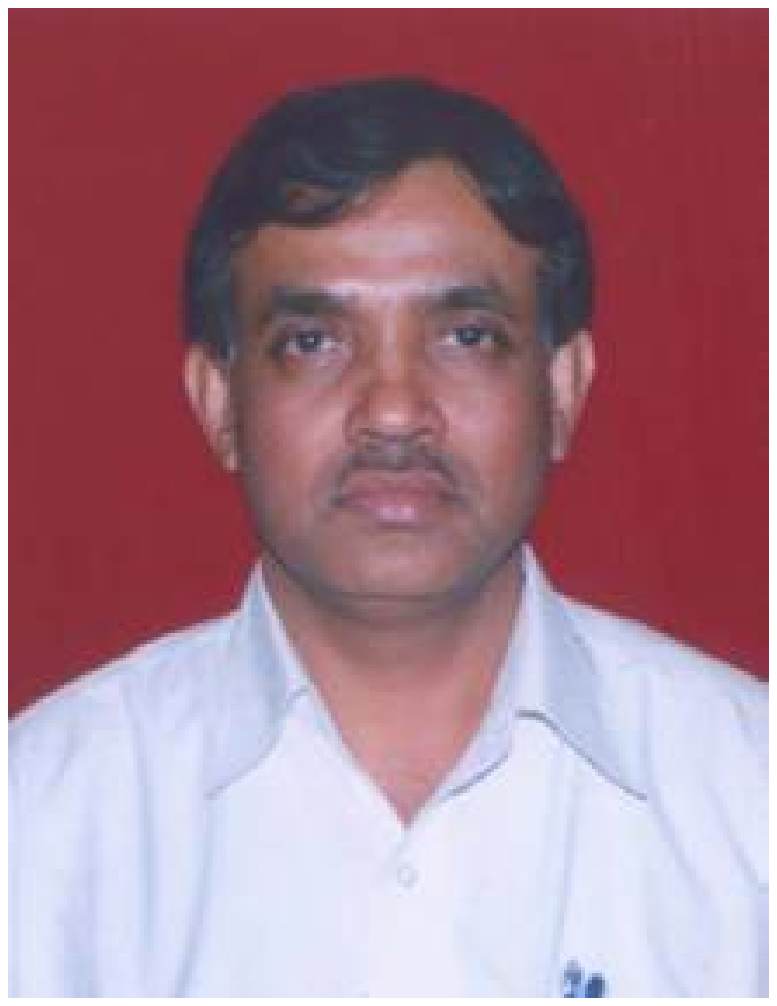}}
\begin{minipage}[b][1in][c]{1.8in}
{\centering{\bf {K R Venugopal}} is currently the Principal, University Visvesvaraya College of Engineering, Bangalore University, Bangalore. He obtained his Bachelor of Engineering from University Visvesvaraya College of Engineering. He received his Masters degree in Computer Science and }\\\\
\end{minipage}
 Automation from Indian Institute of Science Bangalore. He was awarded Ph.D in Economics from Bangalore University and Ph.D in Computer Science from Indian Institute of Technology, Madras. He has a distinguished academic career and has degrees in Electronics, Economics, Law, Business Finance, Public Relations, Communications, Industrial Relations, Computer Science and Journalism. He has authored 39 books on Computer Science and Economics, which include Petrodollar and the World Economy, C Aptitude, Mastering C, Microprocessor Programming, Mastering C++ and Digital Circuits and Systems $etc.$. During his three decades of service at UVCE he has over 350 research papers to his credit. His research interests include Computer Networks, Wireless Sensor Networks, Parallel and Distributed Systems, Digital Signal Processing and Data Mining. \\\\\\
\noindent{\includegraphics[width=1in,height=1.7in,clip,keepaspectratio]{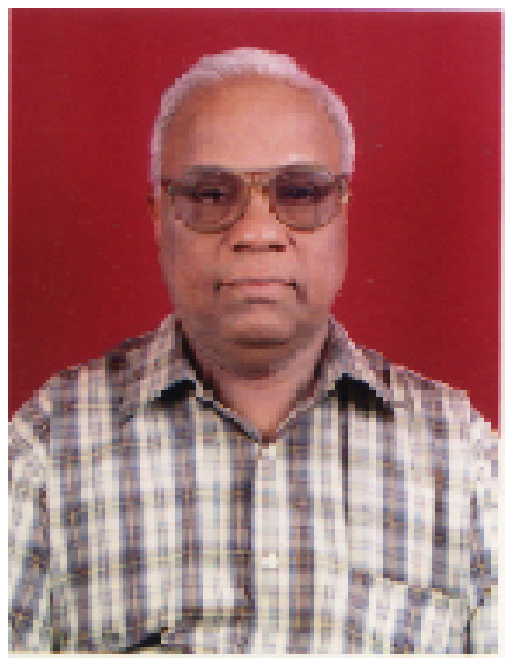}}
\begin{minipage}[b][1in][c]{1.8in}
{\centering{\bf{L M Patnaik }} is currently Honorary Professor, Indian Institute of Science, Bangalore, India. He was a Vice Chancellor, Defense Institute of Advanced Technology, Pune, India and was a Professor since 1986 with the Department of Computer Science and Automation, Indian  } \\\\
\end{minipage}
Institute of Science, Bangalore. During the past 35 years of his service at the Institute he has over 500 research publications in refereed International Journals and Conference Proceedings. He is a Fellow of all the four leading Science and Engineering Academies in India; Fellow of the IEEE and the Academy of Science for the Developing World. He has received twenty national and international awards; notable among them is the IEEE Technical Achievement Award for his significant contributions to High Performance Computing and Soft Computing. His areas of research interest have been Parallel and Distributed Computing, Mobile Computing, CAD for VLSI circuits, Soft Computing and Computational Neuroscience.\\\\
\balance
\end{document}